\documentclass[12pt]{article}
\usepackage{eqsection,latexsym,epsf,epsfig,cite}

\begin{document} 

\setlength{\unitlength}{0.2cm}

\title{
Structure factor of dilute ring polymers
}
\author{
  \\
  {\small Pasquale Calabrese}             \\[-0.2cm]
  {\small\it 
           Scuola Normale Superiore and INFN -- Sezione di Pisa}  \\[-0.2cm]
  {\small\it I-56100 Pisa, ITALY}         \\[-0.2cm]
  {\small Internet: {\tt Pasquale.Calabrese@df.unipi.it}}     \\[-0.2cm]
        \\[-0.1cm]  
  \\[-0.1cm]  \and
  {\small Andrea Pelissetto}                          \\[-0.2cm]
  {\small\it Dipartimento di Fisica and INFN -- Sezione di Roma I} \\[-0.2cm]
  {\small\it Universit\`a degli Studi di Roma ``La Sapienza"}        \\[-0.2cm]
  {\small\it I-00185 Roma, ITALY}          \\[-0.2cm]
  {\small Internet: {\tt Andrea.Pelissetto@roma1.infn.it}}   \\[-0.2cm]
        \\[-0.1cm]  
  \\[-0.1cm]  \and
  {\small Ettore Vicari}                          \\[-0.2cm]
  {\small\it Dipartimento di Fisica and INFN -- Sezione di Pisa} \\[-0.2cm]
  {\small\it Universit\`a degli Studi di Pisa}        \\[-0.2cm]
  {\small\it I-56100 Pisa, ITALY}          \\[-0.2cm]
  {\small Internet: {\tt Ettore.Vicari@df.unipi.it}}   \\[-0.2cm]
  {\protect\makebox[5in]{\quad}}  
  \\
}
\vspace{0.5cm}

\maketitle
\thispagestyle{empty}   

\vspace{0.2cm}

\begin{abstract}
We consider ring polymers in good solvents in the dilute limit. 
We determine the structure factor and the 
monomer-monomer distribution function. We compute accurately the asymptotic 
behavior of these functions for small and large momenta and distances
by using field-theoretical methods. Phenomenological expressions
with the correct asymptotic behaviors are also given.
\end{abstract}

\clearpage

\newcommand{\be}{\begin{equation}}
\newcommand{\ee}{\end{equation}}
\newcommand{\beq}{\begin{equation}}
\newcommand{\eeq}{\end{equation}}
\newcommand{\bea}{\begin{eqnarray}}
\newcommand{\eea}{\end{eqnarray}}
\newcommand{\<}{\langle}
\renewcommand{\>}{\rangle}

\def\spose#1{\hbox to 0pt{#1\hss}}
\def\ltapprox{\mathrel{\spose{\lower 3pt\hbox{$\mathchar"218$}}
 \raise 2.0pt\hbox{$\mathchar"13C$}}}
\def\gtapprox{\mathrel{\spose{\lower 3pt\hbox{$\mathchar"218$}}
 \raise 2.0pt\hbox{$\mathchar"13E$}}}
\def\lle{<<}
\def\gge{>>}

\newcommand{\R}{\hbox{{\rm I}\kern-.2em\hbox{\rm R}}}
\def\bsigma{\mbox{\protect\boldmath $\sigma$}}

\newcommand{\reff}[1]{(\ref{#1})}
\def\smfrac#1#2{{\textstyle\frac{#1}{#2}}}

\section{Introduction}

The statistical properties of dilute polymers in good solvents have been
the subject of extensive studies during the years 
\cite{Flory_book,DeGennes_book,DesCloizeaux-Jannink_book,Freed_book,Oono}. 
A significant understanding
of the problem was reached when it was realized that long polymers 
could be modelled by chains with an excluded-volume
interaction. This allowed the introduction of simplified
theoretical models which could be analyzed more easily.
{}From a theoretical point of view, an important step forward was
made by de Gennes \cite{DeGennes_72}, 
who proved that the statistical properties 
of dilute polymers could be obtained as the limit $n\to 0$ of the 
$n$-component $\phi^4$ theory, opening the field to the many
methods that have been developed for the study 
of the critical behavior of spin systems.

In nature polymers may have many different geometrical conformations.
In this paper we will focus on ring polymers and we will determine 
the structure factor and the monomer-monomer distribution function.
For this purpose we will use the mapping with the 
$\lambda\phi^4$ theory \cite{DeGennes_72,Daoud,DC,Emery,ACF} 
and some recent theoretical results for the two-point function of 
biquadratic operators \cite{NA-97,CPV-01}. Indeed, we show that 
the structure factor is directly related to the 
energy-energy correlation function. Then,
by extending the results available for the end-to-end distribution function
\cite{Fisher-Hiley_61,Fisher_66,Mazur_65,McKenzie-Moore_71,%
McKenzie_76,DesCloizeaux_74_80,CCP-00,footnote0}, 
we compute its large- and small-momentum behavior.
Moreover, by using an interpolation formula based on a dispersive 
approach \cite{Bray-76,CPV-01}, we obtain a general expression 
valid for all momenta, i.e. for all values of the radiation (neutron)
wavelength.  We also investigate the 
monomer-monomer distribution function that is closely related to the 
structure factor by a Fourier transform. Its properties for small and large 
distances are investigated in detail.

The paper is organized as follows. 
In Sec. \ref{sec2} we define the structure factor and the monomer-monomer
distribution function for $N$-step ring polymers. 
In Sec. \ref{sec3.1} we show that the computation of the structure 
factor is equivalent to the calculation of the energy-energy correlation
function in the $n$-component $\lambda\phi^4$ theory in the limit $n\to 0$. 
Field-theoretical results for such a correlation function are reviewed 
in Sec. \ref{sec3.2}. In Sec. \ref{sec4} we study the structure factor.
In particular, we determine its small-$q^2$ behavior extending the 
classical Guinier formula, its large-$q^2$ behavior, and give 
a general interpolation formula that has the correct asymptotic behavior 
for $q^2\to 0$ and $q^2\to\infty$. Finally, in Sec. \ref{sec5} 
we report the corresponding expressions for the monomer-monomer distribution
function. We also verify that the phenomenological expression 
often used for the end-to-end distribution function for linear polymers
\cite{Mazur_65,McKenzie-Moore_71,McKenzie_76,DesCloizeaux_74_80} 
also provides a good approximation to the monomer-monomer distribution
function considered here.

\section{Definitions} \label{sec2}

We consider a monodisperse ensemble of ring polymers with $N$ monomers, 
labelling the monomers from 1 to $N$. Then, we consider two monomers 
$i$ and $j$ of the walk and the unnormalized distribution $d_{ij,N}(\vec{r})$
of the distance $\vec{r}$ between the two monomers.
Such a distribution function depends only 
on the relative position $|i-j|$ of the monomers along the polymer. 
Then, we define its average over all pairs $i,j$, i.e.
\begin{equation}
p_N(\vec{r}) = {1\over N^2} \sum_{i,j=1}^N d_{ij,N}(\vec{r}),
\end{equation}
the corresponding normalized distribution 
\begin{equation}
P_N(\vec{r}) = {p_N(\vec{r})\over \int d^d \vec{s}\, p_N(\vec{s})},
\end{equation} 
and the mean squared radius of gyration
\begin{equation} 
R^2_{g,N} = {1\over 2} \int d^d \vec{s}\, |s|^2 P_N(\vec{s}).
\label{defR2gN}
\end{equation}
On a lattice, 
$d_{ij,N}(\vec{r})$ can be identified with the number of $N$-step lattice 
rooted self-avoiding polygons such that monomer $i$ is in the origin and 
monomer $j$ is in $\vec{r}$. 
Note that 
\begin{equation}
p_N\equiv \sum_s d_{ij,N}(\vec{s}) = \sum_s p_N(\vec{s})
\end{equation}
is the total number of $N$-step lattice rooted polygons.

The structure factor is defined by
\begin{equation} 
S_N(\vec{q}) = \int d^d \vec{s}\, e^{i\vec{q}\cdot\vec{s}} P_N(\vec{s}),
\label{defSN}
\end{equation}
which, by definition, satisfies $S_N(0) = 1$. For elastic scattering, 
the momentum $\vec{q}$ is 
directly related to the wavelength $\lambda$ of the incoming radiation 
in the scattering medium and to the scattering angle $\theta$ by 
\begin{equation}
|q| = {4 \pi\over \lambda} \sin {\theta\over 2}.
\end{equation}

For comparison, we will also consider linear polymers and 
correspondingly we define 
the mean squared end-to-end distance $\widehat{R}^2_{e,N}$ and the 
mean squared radius of gyration $\widehat{R}^2_{g,N}$. 
In the limit $N\to \infty$,
the radius of gyration and the end-to-end distance diverge with the same 
critical exponent $\nu$, i.e.
\begin{eqnarray}
R^2_{g,N} &=&  a_g N^{2\nu}, \nonumber \\
\widehat{R}^2_{g,N} &=&  \hat{a}_g N^{2\nu}, \nonumber \\
\widehat{R}^2_{e,N} &=&  \hat{a}_e N^{2\nu}. 
\label{Rasymptotic}
\end{eqnarray}
In two dimensions the universal exponent $\nu$ is given by 
$\nu = 3/4$, while in three dimensions 
\begin{equation}
\nu = \cases{0.5877(6) & \hskip 1truecm Ref. \cite{Li-etal_95} \cr
             0.58758(7) & \hskip 1truecm Ref. \cite{Belohorec-Nickel_97}.
            }
\end{equation}
See Ref. \cite{review} for other numerical and theoretical estimates.

In the limit $N\to\infty$, the normalized distribution $P_N(\vec{r})$ and the 
structure factor $S_N(\vec{q})$ obey general scaling laws. More precisely,
for $N\to\infty$, $|r|\to\infty$, $|q|\to 0$, 
with $\vec{\rho} \equiv \vec{r}/R_{g,N}$
and $\vec{Q} \equiv q R_{g,N}$ fixed, we have
\begin{eqnarray} 
P_N(\vec{r}) &\approx& {1\over R^d_{g,N}} f(\rho), \nonumber \\
S_N(\vec{q})  &\approx& s(Q) .
\label{scaling-SAW}
\end{eqnarray}
These functions satisfy the following normalization conditions:
\begin{eqnarray} 
\int d\vec{\rho}\, f(\rho) = 1 \qquad && \qquad 
\int d\vec{\rho}\, \rho^2 f(\rho) = 2,
\nonumber \\
s(0) = 1 \qquad && \qquad 
\left. {ds(Q)\over dQ^2}\right|_{Q = 0} = - {1\over 3}.
\label{normalizations}
\end{eqnarray}

\section{Field-theoretical results} \label{sec3}

\subsection{Generalities} \label{sec3.1}

We wish now to derive some general properties of the functions we have 
introduced above. We will use the Laplace-de Gennes transform method 
\cite{DeGennes_72,DesCloizeaux_74_80,DesCloizeaux-Jannink_book}
and the results 
for biquadratic correlation functions in $n$-vector models
obtained in Refs.~\cite{NA-97,CPV-01}. 

For this purpose, consider a hypercubic lattice in $d$ dimensions and the 
$n$-vector model Hamiltonian 
\begin{equation}
H = - \beta \sum_{\langle rs\rangle} \mbox{\protect\boldmath $\sigma$}_r\cdot 
                               \mbox{\protect\boldmath $\sigma$}_s,
\end{equation}
where ${\bsigma}_r$ is a unit $n$-dimensional vector and the sum 
is extended over all lattice nearest-neighbor sites $\langle rs\rangle$. 
Then, consider the correlation function 
\begin{equation}
C_\mu(\vec{r};\beta) \equiv
\langle ({\bsigma}_0\cdot {\bsigma}_\mu) 
        ({\bsigma}_r\cdot {\bsigma}_{r+\mu}) \rangle - 
        \langle ({\bsigma}_0\cdot {\bsigma}_\mu)\rangle^2,
\end{equation} 
where $\vec{\mu}$ is an arbitrary lattice unit vector. 
By using standard results 
(see, e.g., Refs. \cite{DeGennes_book,DesCloizeaux-Jannink_book}),
it is easy to verify that 
\begin{equation}
\lim_{n\to 0} \left[{1\over n} C_\mu(\vec{r};\beta) \right] =\, 
  \sum_N \beta^{N-2} c_{N,\mu} (\vec{r}),
\end{equation}  
where $c_{N,\mu} (\vec{r})$ is the number of $N$-step
lattice self-avoiding polygons that 
go through the lattice links $\langle 0,\vec{\mu}\rangle$ and 
$\langle \vec{r},\vec{r}+\vec{\mu}\rangle$. 
In the limit $N\to\infty$, $|r|\to\infty$ 
we are interested in, $c_{N,\mu} (\vec{r})$ is proportional to the number of 
lattice polygons $p_N(\vec{r})$ going through $0$ and $\vec{r}$. 
Then
\begin{equation}
\lim_{n\to 0} \left[{1\over n} C_\mu(\vec{r};\beta) \right] \approx\, 
{\rm const} \times \sum_N \beta^N p_N P_N(\vec{r}).
\end{equation}
Now, let us consider the left-hand side. 
In the critical limit it can be identified 
with the energy-energy correlation function in the $\lambda \phi^4$ theory.
More precisely, if $t_\sigma \equiv (\beta_c - \beta)/\beta_c \to 0$ is 
the reduced temperature in the $n$-vector model, we have
\begin{equation}
C_\mu(\vec{r};\beta) \approx {\rm const}\times G_E(\vec{r},k t_\sigma),
\end{equation}
where $k$ is a constant and 
\begin{equation} 
G_E(\vec{r},t) \equiv
  \langle \phi^2(0) \phi^2(\vec{r})\rangle_t - \langle \phi^2(0)\rangle_t^2,
\end{equation}
where $\langle \cdot \rangle_t$ indicates that the average should be taken at 
reduced temperature $t$ in the $\lambda \phi^4$ theory. 

For $N\to\infty$ one can then derive
(see, e.g., Ref. \cite{DesCloizeaux-Jannink_book})
\begin{equation}
P_N(\vec{r}) \approx {\rm const}\times 
  \int^{+i\infty}_{-i\infty} {dt\over 2\pi i}\, 
   e^{Nt}\,  \lim_{n\to 0} \left[{1\over n} G_E(\vec{r},kt)\right],
\label{PNLaplace}
\end{equation}
where the integral is over a line parallel to the imaginary axis that leaves 
all singularities in the left-hand side.

\subsection{Behavior of $G_E(\vec{r},t)$} \label{sec3.2}

\subsubsection{General results} \label{sec3.2.1}

The general properties of $G_E(\vec{r},t)$ and of its Fourier transform 
$\widetilde{G}_E(\vec{q},t)$ were studied in Refs. \cite{NA-97,CPV-01}. 
We will review them
here. In the scaling limit $t\to 0^+$, we have
\begin{equation}
\widetilde{G}_E(\vec{q},t) = 
   A_E^+ t^{-\alpha} f_E(|q| \xi) \left[1 + O(t^\alpha,t^\Delta)\right],
\label{GE-scaling}
\end{equation}
where $\alpha = 2 - d\nu$ ($\alpha =0.23726(21)$ 
in three dimensions), $f_E(y)$ is a universal function, 
and $\xi$ is the second-moment correlation length defined 
in terms of the two-point function of the fundamental field 
$\widetilde{G}_\phi(\vec{q},t)$,
\begin{equation}
\xi^2 = - \widetilde{G}_\phi(0,t)^{-1} \left. 
     {\partial \widetilde{G}_\phi(\vec{q},t)\over \partial q^2} 
         \right|_{q^2 = 0}.
\end{equation}
Notice the presence of two types of corrections in Eq. (\ref{GE-scaling}). 
First of all, there are the corrections due to the leading irrelevant operator 
with exponent $\Delta$.  
Additionally, there are corrections with exponent $\alpha$ due to the 
presence of a background term. Note that both in two and three dimensions
\cite{footnote1} $\alpha < \Delta$, and thus the leading correction 
is due to the background term. 

The function $f_E(y)$ has a regular expansion for $y\to 0$, 
\begin{equation}
f_E(y) = 1 + \sum_{n=1}^\infty e_n y^{2n},
\end{equation}
while for $y\to \infty$ it has the nonanalytic behavior
\begin{equation}
f_E(y) \approx E_1 y^{-\alpha/\nu} 
   \left[1 + E_2 y^{-(1-\alpha)/\nu} + E_3 y^{-1/\nu}\right].
\label{fE-largeq2}
\end{equation}
For generic values of $y$, the universal function $f_E(y)$ can be 
written in the form
\begin{equation} 
f_E(y) = 1 - {y^2 E_1\over \pi} \sin\left({\pi \alpha \over 2\nu}\right)
\int_{4 S^+_M}^\infty dx\, {x^{-1-\alpha/(2\nu)}\over x + y^2} F_E(x),
\label{fE-dispersive}
\end{equation}
where $E_1$ is the coefficient appearing in Eq. (\ref{fE-largeq2}),
$S^+_M$ is a constant, and $F_E(y)$ is the spectral function. 
The constant $S^+_M$ is universal and it is defined by 
$S_M^+ \equiv \xi^2/\xi_{\rm gap}^2$, where $\xi_{\rm gap}$ is the exponential 
correlation length that determines the large-distance behavior of the 
two-point function $G_\phi(\vec{r},t)$ of the 
field $\phi$. The constant $E_1$ and the spectral function are 
related by the sum rule
\begin{equation}
{E_1\over \pi} \sin\left({\pi\alpha\over 2\nu}\right)
\int_{4 S^+_M}^\infty dx\, x^{-1-\alpha/(2\nu)}\ F_E(x)=\, 1.
\label{sumrul-pos}
\end{equation}  
The representation (\ref{fE-dispersive}) is exact under rather mild 
assumptions \cite{CPV-01,Bray-76}. 
Approximate expressions can be obtained by choosing approximate forms for 
the spectral function. As in Ref. \cite{CPV-01}, we choose
\begin{equation}
F_E(x) = 1 + E_2 \Phi_2 x^{-(1-\alpha)/(2\nu)} + E_3 \Phi_3 x^{-1/(2\nu)},
\label{FFL}
\end{equation}
where $E_2$ and $E_3$ are the constants that parametrize the large-$q^2$ 
behavior in Eq. (\ref{fE-largeq2}) and
\begin{eqnarray}
\Phi_2 &=& 
   \cos {\pi(1-\alpha)\over 2\nu} + \sin{\pi(1-\alpha)\over 2\nu}
         \cot {\pi\alpha\over 2\nu},
\nonumber \\
\Phi_3 &=& 
   \cos {\pi\over 2\nu} + \sin{\pi\over 2\nu}
         \cot {\pi\alpha\over 2\nu}.
\end{eqnarray}                                                                  
We further define
\begin{eqnarray}
g(p;x) &\equiv& - {\pi\over \sin\left({\pi\alpha\over 2\nu} + \pi p\right)} \, 
            x^{-1-p-\alpha/(2\nu)} 
\nonumber \\
&& \qquad + {2\nu\left(4 S_M^+\right)^{-p-\alpha/(2\nu)} \over 
          (\alpha + 2 p \nu) (x + 4 S_M^+)} \, 
       {}_2F_1 \left(1,1;1-p-{\alpha\over2\nu};
           {4 S_M^+\over x + 4 S_M^+}\right),
\end{eqnarray}
where ${}_2F_1(a,b;c;z)$ is a hypergeometric function \cite{Gradshtein}.
Then
\begin{equation}
f_E(y) = 1 - {y^2 E_1\over \pi} \sin\left({\pi\alpha\over 2\nu}\right)
   \left[g(0;y^2) + E_2 \Phi_2 g\left({1-\alpha\over2\nu};y^2\right)
                  + E_3 \Phi_3 g\left({1\over2\nu};y^2\right)\right],
\label{fEFL1}
\end{equation}
or, by using the sum rule (\ref{sumrul-pos}), 
\begin{equation}
f_E(y) = {E_1\over \pi} \sin\left({\pi\alpha\over 2\nu}\right)
   \left[g(-1;y^2) + E_2 \Phi_2 g\left(-1+{1-\alpha\over2\nu};y^2\right)
                  + E_3 \Phi_3 g\left(-1+{1\over2\nu};y^2\right)\right].
\label{fEFL2}
\end{equation}
{}From Eq. (\ref{GE-scaling}) we also obtain 
\begin{equation}
G_E(\vec{r},t) = {A^+_E t^{-\alpha}\over \xi^d} \widetilde{f}_E (s) 
\left[ 1 + O(t^\alpha, t^\Delta)\right],
\end{equation}
where $\vec{s} \equiv \vec{r}/\xi$, and 
\begin{equation} 
\widetilde{f}_E (s) = 
    \int {d^d \vec{y}\over (2\pi)^d}\,  f_E(y) e^{i\vec{s}\cdot \vec{y}}.
\label{deftildefE}
\end{equation}
For $s\to\infty$, $\widetilde{f}_E (s)$ decays exponentially as 
\begin{equation}
\widetilde{f}_E (s) = 
    A s^p \exp\left(- {s \xi\over \xi_{E,\rm gap}}\right),
\label{tildefElarger}
\end{equation}
where 
\begin{equation}
\xi_{E,\rm gap} = {1\over 2} \xi_{\rm gap} = {\xi\over 2 \sqrt{S_M^+}},
\end{equation} 
and $p$ is an exponent that can in principle be computed perturbatively.

An approximate expression can be obtained from Eq. (\ref{fEFL2}):
\begin{equation}
\widetilde{f}_E(s) = 
{E_1\over 4 \pi^2 s} \sin\left({\pi\alpha\over 2\nu}\right)
   \left[h(0;s) + E_2 \Phi_2 h\left({1-\alpha\over2\nu};s\right)
                  + E_3 \Phi_3 h\left({1\over2\nu};s\right)\right],
\label{ftildeEFL}
\end{equation}
where 
\begin{equation}
h(p;s) = 2 s^{2p-2+\alpha/\nu} \Gamma\left(2 - 2p -\alpha/\nu,
          2 s\sqrt{S_M^+}\right),
\end{equation} 
and $\Gamma(\alpha;x)$ is an incomplete $\Gamma$-function \cite{Gradshtein}.

\subsubsection{Numerical results} \label{sec3.2.2}

We have introduced in the preceding section several constants that will be 
computed here. 

First, let us consider the constants $e_n$ that parametrize the 
small-momentum expansion
of $f_E(y)$. Perturbative series were derived in Ref. \cite{CPV-01} for 
$e_n$, $n=1,\ldots,5$, in the fixed-dimension expansion in three dimensions 
(four loops) and in $\epsilon$-expansion (three loops). 
We resummed the perturbative series by 
using their large-order behavior and performing 
a conformal mapping as in Ref. \cite{francesi1980}. Mean values ad errors
were computed using the algorithm of Ref.~\cite{CPV-00}. 
For the expansion in fixed dimension we used 
for the four-point renormalized coupling\cite{footnote2}
$\bar{g}^* = 1.395(15)$ that includes all available estimates 
\cite{Guida-ZinnJustin_98,PV-98,PV-00}.
The results are reported in Table \ref{stime-ei}. Note that the 
two expansions give estimates that agree within error bars, confirming the 
correctness of our results  within the quoted errors.

\begin{table}[tbp]
\caption{Estimates of the small-momentum expansion coefficients by using the 
fixed-dimension expansion in $d=3$ ($d=3$) and the $\epsilon$ expansion 
($\epsilon$-exp). ``final" labels our subjective final estimates.
Perturbative series from Ref. \cite{CPV-01}. }
\label{stime-ei}
\begin{tabular}{clllll}
\hline\hline
& 
\multicolumn{1}{c}{$e_1$}&
\multicolumn{1}{c}{$e_2$}&
\multicolumn{1}{c}{$e_3$}&
\multicolumn{1}{c}{$e_4$}&
\multicolumn{1}{c}{$e_5$}
\\ 
\hline
$(d=3)$ & $-$0.0322(11) & 0.00382(15) & $-$0.000597(23) & 
          1.06(4)$\times 10^{-4}$ & $-$2.03(8)$\times 10^{-5}$ \\
($\epsilon$-exp)
        & $-$0.0323(7) & 0.00398(11) & $-$0.000636(10) & 
          1.14(4)$\times 10^{-4}$ & $-$2.22(3)$\times 10^{-5}$ \\
final   & $-$0.0323(10)  & 0.00390(15) & $-$0.000620(30) &
          1.10(8)$\times 10^{-4}$ & $-$2.15(15)$\times 10^{-5}$ \\
\hline\hline
\end{tabular}
\end{table}    

We can also use the results of Ref. \cite{CPV-01} to compute the 
large-momentum expansion coefficients $E_i$. In $\epsilon\equiv 4 - d$
expansion, they are explicitly given by
\begin{eqnarray}
E_1 &=& 1 + {\epsilon\over 2} + O(\epsilon^2), \nonumber  \\
E_2 &=& -2 + {5 \epsilon\over 4} + O(\epsilon^2), \nonumber \\
E_3 &=& 2 - {7 \epsilon\over 4} + O(\epsilon^2).
\end{eqnarray}
In the following we will also
be interested in the product $E_1 E_2$ given by
\begin{eqnarray}
E_1 E_2 = -2 + {\epsilon\over4} + O(\epsilon^2).
\end{eqnarray} 
The large size of the coefficients makes it difficult to resum the 
perturbative series. We report here the results obtained by setting 
$\epsilon = 1$ and as error we quote the last coefficient:
$E_1 = 1.5(5)$, $E_2 = -0.75(1.25)$, $E_3 = 0.25 (1.75)$,
$E_1 E_2 = -1.75(25)$.
For the sum $E_2 + E_3$ an additional term is known:
\begin{equation}
E_2 + E_3 = -{1\over2}\epsilon +\left({27\over64} - {\pi^2\over24}\right)
   \epsilon^2 + O(\epsilon^3),
\end{equation}
so that $E_2 + E_3 = - 0.49(1)$.

We also need the universal ratio $S_M^+$. In three dimensions it has 
been estimated by a variety of field-theoretical and exact-enumeration 
methods \cite{CPRV-98}, obtaining 
$S_M^+ - 1 = - 3(1) \times 10^{-4}$. In two dimensions, an exact-enumeration 
study gives \cite{Campostrini-etal_96} $S_M^+ - 1 = 1(2) \times 10^{-4}$.

We wish finally to determine the functions $f_E(y)$ and 
$\tilde{f}_E(\rho)$ by using Eqs. (\ref{fEFL2}) and 
(\ref{ftildeEFL}). For this purpose, we must fix $\alpha$, $\nu$,
$S_M^+$, $E_2$, and $E_3$, while $E_1$ is fixed by using the sum rule 
(\ref{sumrul-pos}). For $\nu$ and $\alpha = 2 - 3 \nu$, we use 
\cite{Belohorec-Nickel_97} $\nu = 0.58758$, while for $S_M^+$ 
we use the above-reported result. For $E_2$ and $E_3$, the only available 
estimates are those obtained in the $\epsilon$-expansion approach.
They have a large error, while, apparently, their sum is more precisely 
determined. We have thus followed the following strategy. We have fixed 
$E_2 + E_3 = -0.49$ using the $\epsilon$-expansion. Then, we have chosen
$E_2$ and $E_3$ so that for small $y$ the approximation gives 
$1 - 0.0323 y^2$, in agreement with the results of Table \ref{stime-ei}. 
In this way we obtain, $E_2 = -1.38$, $E_3 = 0.89$, and $E_1 = 1.60$. These 
estimates are in good agreement with the $\epsilon$-expansion 
results, confirming that our approximation is reasonably correct 
for $y\to \infty$. Moreover, it also nicely reproduces the 
small $y$-behavior. Indeed, it gives $e_2 \approx 0.0040$, 
$e_3 \approx -0.00065$, $e_4 \approx 1.2\times 10^{-4}$, 
which are close to the results of Table \ref{stime-ei}.

\section{The structure factor $S_N(\vec{q})$} \label{sec4}

Using Eqs. (\ref{defSN}), (\ref{PNLaplace}), and 
(\ref{GE-scaling}) we can determine the structure factor 
$S_N(\vec{q})$ in the scaling limit $|q|\to 0$, $N\to \infty$, 
with $Q\equiv |q| R_{N,g}$ fixed. Neglecting corrections of order 
$N^{-\alpha}$, see the discussion in Sec. \ref{sec3.2.1}, we 
can write---from now on, 
the limit $n\to 0$ is always understood---
\begin{equation}
S_N(\vec{q}) = \Gamma(\alpha) 
   \int_{-i\infty}^{+i\infty} {dt\over 2\pi i}\, e^t t^{-\alpha} 
        f_E\left(|q| N^{\nu} x_0 k^{-\nu} t^{-\nu}\right),
\label{SNLaplace}
\end{equation}
where we have written for the second-moment correlation length $\xi$,
$\xi\approx x_0 t^{-\nu}$ for $t\to 0$. The nonuniversal factor can be 
eliminated by introducing the end-to-end distance for linear 
polymers. A simple calculation gives
\cite{DesCloizeaux-Jannink_book,DesCloizeaux_74_80} 
\begin{equation}
\widehat{R}^2_{e,N} = 2d {\Gamma(\gamma)\over \Gamma(\gamma + 2\nu)} 
    x_0^2 k^{-2\nu} N^{2\nu},
\end{equation}
where $\gamma$ is a universal critical exponent.
If we define 
\begin{equation}
\kappa \equiv {1\over 2d} {\Gamma(\gamma + 2\nu)\over \Gamma(\gamma)},
\end{equation}
then
\begin{equation}
S_N(\vec{q}) = \Gamma(\alpha)
   \int_{-i\infty}^{+i\infty} {dt\over 2\pi i} e^t t^{-\alpha}
   f_E\left(\sqrt{\kappa}\ |q| \widehat{R}_{e,N} t^{-\nu}\right).
\end{equation}
We can now use the results of the preceding section. 
First, for $|q|\to 0$ we obtain 
\begin{equation}
S_N(\vec{q}) = 1 + \sum_{n=1}^\infty e_n 
    {\Gamma(\alpha)\over \Gamma(\alpha + 2 n \nu)} 
    \left(\kappa q^2 \widehat{R}^2_{e,N}\right)^n.
\label{SN-espansione}
\end{equation}
In order to express $S_N(\vec{q})$ in the scaling form (\ref{scaling-SAW}), 
we must express $\widehat{R}^2_{e,N}$ in terms of $R_{g,N}^2$. Using 
Eqs. (\ref{defR2gN}) and (\ref{SN-espansione}) we have in 
$d$ dimensions
\begin{equation}
R_{g,N}^2 = - d \left. {dS_N(\vec{q})\over dq^2} \right|_{q^2=0} = 
     - d e_1 {\Gamma(\alpha)\over \Gamma(\alpha + 2 \nu)} 
       \kappa \widehat{R}^2_{e,N},
\end{equation}
which allows to determine the ratio \cite{footnote3}
\begin{equation} 
H\equiv {R_{g,N}^2 \over \widehat{R}^2_{e,N}} = 
  - d e_1\kappa {\Gamma(\alpha)\over \Gamma(\alpha + 2 \nu)} \; .
\end{equation}
Setting 
\begin{equation}
\bar{\kappa} = - {1\over d e_1} {\Gamma(\alpha + 2\nu)\over \Gamma(\alpha)},
\end{equation}
we have 
\begin{equation}
S_N(\vec{q}) = 1 + \sum_{n=1}^\infty e_n
    {\Gamma(\alpha)\over \Gamma(\alpha + 2 n \nu)}
     \bar{\kappa}^n Q^{2n}\; .
\label{Sq-smallQ}
\end{equation}
In three dimension ($d=3$), using \cite{Belohorec-Nickel_97}
$\nu = 0.58758(7)$, $\alpha = 0.23726(21)$, and 
\cite{Caracciolo-etal_97} $\gamma = 1.1575(6)$, 
we obtain
\begin{equation}
\kappa = 0.21315(12) , \qquad\qquad \bar{\kappa} = 2.40(8).
\label{valorikappa-3d}
\end{equation}
Thus 
\begin{equation} 
H = 0.089(3),
\label{ratioRG-Rhe}
\end{equation} 
and 
\begin{eqnarray}
S_N(\vec{q}) &=& 1 - {1\over3} Q^2 + 0.061(5)\times Q^4 - 0.0073(7) \times Q^6 
  \nonumber \\ 
  && \quad + 
         0.00063(10)\times Q^8 - 0.000045(8)\times Q^{10} + O(Q^{12}).
\label{SN-smallQ2}
\end{eqnarray}
Note that the coefficients decrease quite rapidly, so that this 
expansion provides a good approximation up to $Q\approx 2$.

We can compare the result (\ref{ratioRG-Rhe}) with the existing 
estimates. The amplitude $\hat{a}_e$, cf. Eq. (\ref{Rasymptotic}),
is quite well determined 
for self-avoiding walks on the cubic lattice: 
Ref. \cite{Li-etal_95} quotes $\hat{a}_e = 1.2167(50)$, 
while Ref. \cite{MJHMJG-00} gives $\hat{a}_e \approx 1.225$. 
The amplitude $a_g$, cf. Eq. (\ref{Rasymptotic}),
for self-avoiding polygons on the cubic 
lattice can be obtained from the data reported in Ref. 
\cite{JEK-92}. Fitting their data for $R^2_{g,N}$ with 
$a_g N^{2 \nu} + b_g N^{2 \nu - \Delta}$ and using the 
estimates of $\nu$ and $\Delta$ of Ref. \cite{Belohorec-Nickel_97},
we obtain $a_g \approx 0.102$. Unfortunately, no error bars are 
reported in Ref. \cite{JEK-92} and thus we cannot quote an error 
on our result. Moreover, the algorithm of Ref. \cite{JEK-92} samples 
only polygons with trivial knot type, and thus some systematic error 
can in principle be present. Using these estimates, we obtain 
$H\approx 0.083$, which is reasonably close to our result
(\ref{ratioRG-Rhe}).

It is interesting to compute $e_1$ and $\bar{\kappa}$ in two dimensions,
by using the available estimates of $H$, 
$H = 0.14605(7)$ (Ref. \cite{CG-93}) and $H = 0.1459(2)$
(Ref. \cite{Lin-00}). 
Since $\nu = 3/4$ and \cite{Nienhuis-82-84} $\gamma = 43/32$, 
we obtain $e_1 = - 0.08447(4)$, $\bar{\kappa} = 3.3398(16)$.

For $Q^2\to \infty$, we can use Eqs. (\ref{SNLaplace}) and (\ref{fE-largeq2}) 
to obtain 
\begin{equation} 
S_N(\vec{q})\approx (\alpha -1) E_1 E_2 
   \left(\kappa q^2 \widehat{R}^2_{e,N}\right)^{-1/(2\nu)} 
   \approx (\alpha -1) E_1 E_2 \bar{\kappa}^{-1/(2\nu)} Q^{-1/\nu}.
\label{SN-largeQ}
\end{equation}
In three dimensions, using Eq. (\ref{valorikappa-3d}) and the estimate of 
$E_1 E_2$ of Sec. \ref{sec3.2.2}, we have
\begin{equation}
S_N(\vec{q}) \approx -0.363(10)\times E_1 E_2 Q^{-1/\nu} \approx 
                     0.63(9) \times  Q^{-1/\nu}.
\end{equation}
Finally, we can determine an approximate form for $S_N(\vec{q})$, by 
using the interpolation formulae (\ref{fEFL1}), (\ref{fEFL2})
derived in Sec. \ref{sec3.2.1}. 
We obtain 
\begin{eqnarray}
S_N(\vec{q}) &=& 1 + (\alpha-1) E_1 E_2 \bar{\kappa}^{-1/2\nu} Q^{-1/\nu}
\label{SN-Bray}
\\
&& - {E_1\over \pi} \sin\left({\pi\alpha\over 2\nu}\right)
 \left[\hat{g}(0;Q^2) + E_2 \Phi_2 \hat{g}\left({1-\alpha\over 2\nu};Q^2\right) 
       + E_3 \Phi_3 \hat{g}\left({1\over 2\nu};Q^2\right) \right],
\nonumber 
\end{eqnarray}
where
\begin{eqnarray} 
\hat{g}(p;Q^2) &=& 
   {2\nu \left(4 S_M^+\right)^{-p-\alpha/(2\nu)} \over \alpha + 2 p \nu} 
  \Gamma(\alpha) \int_{-i \infty}^{+i\infty} {dt\over 2\pi i}\, e^t \, 
   {\bar{\kappa} Q^2 t^{-\alpha-2\nu}\over 
    4 S^+_M + \bar{\kappa} Q^2 t^{-2\nu}} \times
\nonumber \\
&& \qquad \times 
    {}_2F_1\left(1,1;1-p-{\alpha\over2\nu};
     {4 S^+_M \over4 S^+_M + \bar{\kappa} Q^2 t^{-2\nu}}\right).
\label{integraleS}
\end{eqnarray}
The evaluation of $S_N(\vec{q})$ requires the evaluation of the 
integral (\ref{integraleS}). Particular care should be given to the 
branch cuts. There are indeed three cuts in the complex $t$-plane:
the negative real axis and the lines $t = a \exp[\pm i \pi/(2\nu)]$, 
where $a$ is real satisfying $a \ge [\bar{\kappa} Q^2/(4 S_M^+)]^{1/(2\nu)}$.
In practice, we have found convenient to integrate over the lines 
$t = a \exp[\pm i \pi/(2\nu)] + b$, where $a$ runs over the positive real axis 
and $b$ is a fixed positive constant that we have taken equal to one.

\begin{figure}
\centering
\epsfig{figure=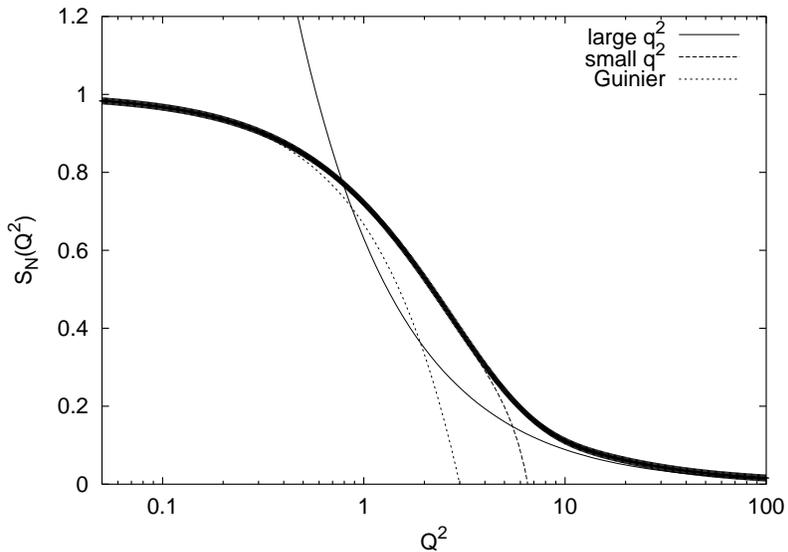,angle=-90,width=0.7\linewidth}
\vspace{0.5cm}
\caption{Structure factor $S_N(\vec{q})$ versus $Q^2 \equiv q^2 R^2_g$. 
We also report the large- and small-$Q^2$ behavior, cf. Eqs. 
(\ref{SN-largeQ}) and (\ref{SN-smallQ2}), and the classical 
Guinier formula $S_N(\vec{q}) = 1 - Q^2/3$.} 
\label{figSN}
\end{figure}                                                                    

\begin{table}[tbp]
\caption{Estimates of the structor factor $S_N(\vec{q})$ for 
$4 \le Q^2 \le 40$, obtained by using the interpolation formula 
(\ref{SN-Bray}).}
\label{tabella-SNq}
\begin{center}
\begin{tabular}{llll}
\hline\hline
\multicolumn{1}{c}{$Q^2$}&
\multicolumn{1}{c}{$S_N(\vec{q})$} &
\multicolumn{1}{c}{$Q^2$}&
\multicolumn{1}{c}{$S_N(\vec{q})$}
\\ 
\hline
4 &   0.305 & 16 & 0.0733 \\
5 &   0.241 & 18 & 0.0667 \\
6 &   0.196 & 20 & 0.0612 \\
7 &   0.164 & 24 & 0.0527 \\
8 &   0.141 & 28 & 0.0464 \\
9 &   0.124 & 30 & 0.0437 \\ 
10 &  0.111 & 34 & 0.0393 \\
12 &  0.093 & 38 & 0.0358 \\ 
14 &  0.082 & 40 & 0.0343 \\
\hline\hline
\end{tabular}
\end{center}
\end{table} 

Using the values of the parameters determined in Sec. \ref{sec3.2.2}, 
we obtain the curve reported in Fig. \ref{figSN}.
We also report the small-$Q^2$ expansion (\ref{SN-smallQ2}) and 
the large-$Q^2$ expansion (\ref{SN-largeQ}).
Note that the small-$Q^2$ approximation is indistinguishible in the graph
from the full curve up to $Q^2 \approx 5$, while the large-$Q^2$ approximation
is reasonably accurate for $Q^2 \gtapprox 40$. Numerical estimates for 
intermediate values are reported in Table \ref{tabella-SNq}.

\begin{figure}
\centering
\epsfig{figure=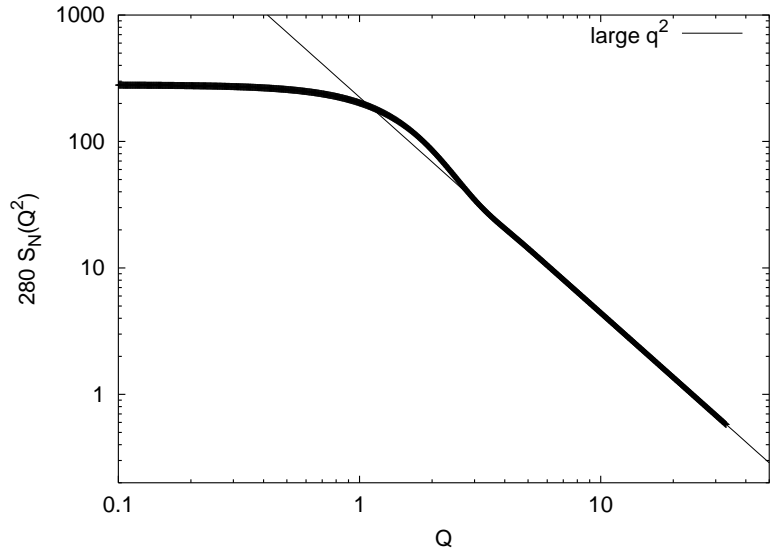,angle=-90,width=0.7\linewidth}
\vspace{0.5cm}
\caption{Structure factor $S_N(\vec{q})$ versus $Q \equiv |q| R_g$. }
\label{SN-JEK}
\end{figure}

We can compare our field-theoretical determination with the 
numerical results of Ref. \cite{JEK-92}. In Fig. \ref{SN-JEK}
we plot $S_N(\vec{q})$ in such a way that it can be directly 
compared with their Fig. 14. We observe a nice quantitative agreement.

\section{Monomer-monomer distribution function $P_N(\vec{r})$} \label{sec5}

We wish now to compute the distribution function $P_N(\vec{r})$, or, 
more precisely, the scaling function $f(\rho)$ defined in 
Eq. (\ref{scaling-SAW}). For this purpose we can either use 
Eq. (\ref{defSN}) and the results of Sec. \ref{sec4} for $S_N(\vec{q})$, 
or use directly the field-theoretical results available for the 
scaling function $\widetilde{f}_E(\rho)$, cf. Eq. (\ref{deftildefE}), and 
\begin{equation}
f(\rho) = \Gamma(\alpha) \bar{\kappa}^{-d/2} 
   \int_{-i\infty}^{i\infty} {dt\over 2\pi i}\, e^t t^{-\alpha+\nu d}\, 
    \widetilde{f}_E\left(\rho t^\nu/\bar{\kappa}^{1/2}\right).
\label{frho-Laplace}
\end{equation}
Using the small-$Q^2$ result (\ref{Sq-smallQ}),
we can compute the moments $M_k$ of $f(\rho)$, i.e.
\begin{equation}
M_k = \int d^d\vec{\rho}\,  |\rho|^{2k} f(\rho).
\end{equation}
A simple computation gives 
\begin{equation}
M_k = (-2\bar{\kappa})^k k! e_k {\Gamma(\alpha)\over \Gamma(\alpha + 2 k\nu)}
   \, \prod_{n=0}^{k-1} (d + 2 n).
\end{equation}
Numerically, in three dimensions, we have 
$M_1 = 2$, $M_2 = 7.3(5)$, $M_3 = 37(4)$, $M_4 = 230(35)$, $M_5 = 1780(330)$.

We can use the large-$q^2$ expansion to compute the small-$\rho$ behavior of 
$f(\rho)$. Explicitly, by using Eq. (\ref{SN-largeQ}), we obtain 
\begin{equation}
f(\rho) \approx p_0 \rho^{1/\nu - d},
\label{frho-smallrho}
\end{equation}
where 
\begin{equation}
p_0 = (\alpha - 1) E_1 E_2 (4 \bar{\kappa})^{-1/(2\nu)} \pi^{-d/2}
    {\Gamma(d/2 - 1/(2 \nu)) \over \Gamma(1/(2 \nu))}.
\end{equation}
Numerically in three dimensions, $p_0 = -0.0250(7) \times E_1 E_2$. 
If we use the 
estimate $E_1 E_2 = -1.75(25)$ reported in Sec. \ref{sec3.2.2} 
we have $p_0 = 0.044(6)$.

Finally, we determine the large-$\rho$ behavior of $f(\rho)$. 
Starting from Eq. (\ref{frho-Laplace}) and 
using Eq. (\ref{tildefElarger}), we have 
\begin{equation} 
f(\rho) \approx \Gamma(\alpha) A \bar{\kappa}^{-d/2-p/2} \rho^p
   \int_{-i\infty}^{+i\infty} {dt\over 2\pi i} e^t t^{-\alpha+\nu d + \nu p} 
   \exp \left( -2 \rho t^\nu \sqrt{S_M^+/\bar{\kappa}} \right).
\end{equation}
The calculation is analogous to that presented in 
Refs. \cite{Fisher_66,DesCloizeaux_74_80}. We obtain that 
$f(\rho)$ obeys the Fisher law 
\begin{equation} 
f(\rho) \sim \rho^q \exp(-D \rho^\delta), 
\end{equation} 
where 
\begin{eqnarray}
\delta &=& {1\over 1 - \nu}, \\
D &=& {1-\nu \over \nu} 
   \left({4 \nu^2 S_M^+\over \bar{\kappa}}\right)^{\delta/2},
\end{eqnarray}
and $q$ is an exponent that can be expressed in terms of $p$, 
cf. Eq. (\ref{tildefElarger}).
Numerically, in three dimensions, $\delta = 2.425(4)$, $D = 0.360(14)$,
while in two dimensions $\delta = 4$, $D = 0.15132(16)$.

Finally, we can use the interpolation formula (\ref{ftildeEFL}) to obtain 
$f(\rho)$ for generic values of $\rho$. We write
\begin{equation}
f(\rho) = {E_1\over 4\pi^2 \rho} \sin\left({\pi\alpha\over 2\nu}\right)
 \left[\hat{h}(0;\rho) + E_2\Phi_2 \hat{h}\left({1-\alpha\over2\nu};\rho\right)
     + E_3 \Phi_3 \hat{h}\left({1\over2\nu};\rho\right)\right],
\end{equation}
where
\begin{equation}
\hat{h}(p;\rho) = {2\Gamma(\alpha)\over\bar{\kappa}} 
  \left(\rho^2 \bar{\kappa}^{-1}\right)^{p-1+\alpha/(2\nu)} 
  \int_{-i\infty}^{+i\infty} {dt\over 2\pi i} \, 
     e^t t^{2p\nu} \Gamma\left(2 - 2p-\alpha/\nu,
           2\rho t^\nu\sqrt{S_M^+/\bar{\kappa}}\right).
\end{equation}
\begin{figure}
\centering
\epsfig{figure=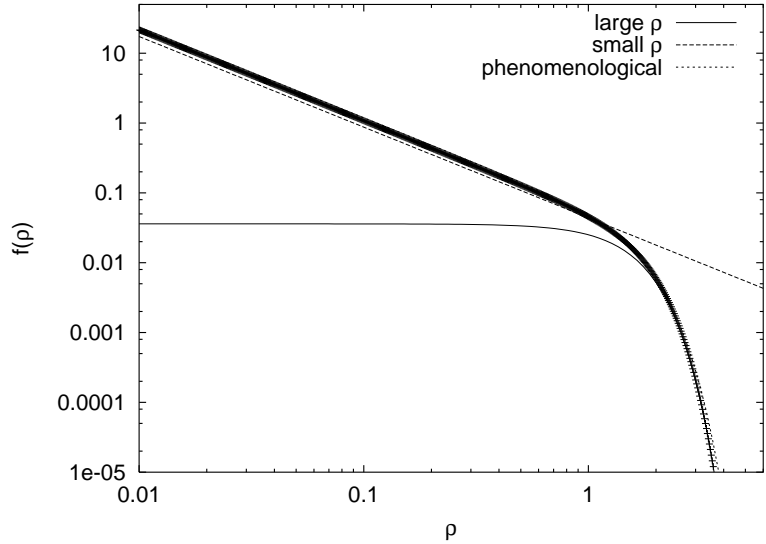,angle=-90,width=0.7\linewidth}
\vspace{0.5cm}
\caption{Monomer-monomer scaling function $f(\rho)$
versus $\rho$. We also report the small-$\rho$ behavior 
(\ref{frho-smallrho}), the curve $0.036 e^{-D\rho^{\delta}}$ 
(labelled ``large $\rho$") that has the correct asymptotic 
behavior for $\rho\to\infty$, and the phenomelogical representation
$0.060 \rho^{1/\nu - 3} e^{-0.273 \rho^\delta}$.}
\label{figfrho}
\end{figure}                                                                    
Using the parameter values determined in Sec. \ref{sec3.2.2} we obtain the 
curve reported in Fig. \ref{figfrho}.

In three dimensions the end-to-end distribution function for linear polymers 
is well described 
by a very simple phenomenological expression
\cite{Mazur_65,McKenzie-Moore_71,McKenzie_76,DesCloizeaux_74_80,CCP-00},
\begin{equation}
f_{\rm ph}(\rho) = A_{\rm ph} \rho^\theta \, e^{- D_{\rm ph} \rho^\delta}\; ,
\end{equation}
where $\delta$ and $\theta$ are fixed by the large-$\rho$ and small-$\rho$
behavior, and the constants $A_{\rm ph}$ and $D_{\rm ph}$ 
by the normalization conditions (\ref{normalizations}):
\begin{eqnarray}
D_{\rm ph} &=& \left[{\Gamma[(1-\nu)(\theta + 5)]\over 
                  2 \Gamma[(1-\nu)(\theta + 3)]} \right]^{\delta/2}, 
\nonumber \\
A_{\rm ph} &=& 
   {\delta \over 4 \pi  \Gamma[(1-\nu)(\theta + 3)]} \,
   D_{\rm ph}^{(1-\nu)(\theta + 3)}.
\end{eqnarray}
An equally good agreement is observed for the function $f(\rho)$ we consider 
here, at least in the relevant region of $\rho$ not too large
($\rho \ltapprox 3$), if we use
$\theta = 1/\nu - 3$, and $D_{\rm ph} = 0.273$, $A_{\rm ph} = 0.060$, 
see Fig. \ref{figfrho}.


\begin{thebibliography}{199}
\bibitem{footnote0} 
The different proposals appearing in the literature and many 
old numerical results have been reviewed by 
Bishop and Clarke \cite{Bishop-Clarke_91}. A list of more recent 
numerical results appears in Ref. \cite{CCP-00}.
\bibitem{footnote1}
In three dimensions
the most precise Monte Carlo estimate of $\Delta$ is
$\Delta = 0.517\pm 0.007 + 0.010$ \cite{Belohorec-Nickel_97}.
Perturbative field theory gives a lower estimate
\cite{Guida-ZinnJustin_98}: 
$\Delta = 0.478\pm 0.010$ ($d=3$ expansion) and 
$\Delta = 0.486 \pm 0.016$ ($\epsilon$-expansion). 
In two dimensions, the value of $\Delta$ is still the object of 
intense debate 
\cite{Guttmann-Enting_88,Saleur_87,Conway-Guttmann_93,%
Conway-Guttmann_96,Guim-etal_97,Caracciolo-etal_98}, 
some models showing $\Delta = 1$, others $\Delta = 11/16$. 
These estimates should be compared with 
\cite{Belohorec-Nickel_97} $\alpha = 0.23726(21)$ 
in three dimensions and $\alpha = 1/2$ in two dimensions.
Thus, in all cases $\alpha < \Delta$.
\bibitem{footnote2} The renormalized  coupling constant $g$ can be expressed in 
terms of the second virial coefficient for a polydisperse ensemble
of polymers, see Ref. \cite{Nickel-Muthukumar}.
\bibitem{footnote3}
This ratio is universal. However, it should be noted that we are 
comparing here the linear and the ring conformation for the 
{\em same} microscopic model, i.e. for polymers of the same 
chemical composition.
\end{thebibliography}
\end{document}